\documentclass[aps,prd,secnumarabic,nobibnotes,twocolumn,superscriptaddress]{revtex4-1}

\usepackage{amsfonts}
\usepackage{mathrsfs}
\usepackage{amsmath}
\usepackage{color}
\usepackage{natbib}
\usepackage{graphicx}
\usepackage{bm}
\usepackage{amssymb}
\usepackage{xspace}
\usepackage{epstopdf}
\usepackage{dcolumn}
\usepackage{longtable}
\usepackage{multirow}
\usepackage[colorlinks=true, letterpaper=true, pdfstartview=FitV, linkcolor=blue, citecolor=blue, urlcolor=blue]{hyperref}

\begin{document}

\title{Goos-H\"{a}nchen-like shifts at metal/superconductor interface}

\author{Ying Liu}\email{ying\_liu@mymail.sutd.edu.sg}
\affiliation{Research Laboratory for Quantum Materials, Singapore University of Technology and Design, Singapore 487372, Singapore}

\author{Zhi-Ming Yu}\email{zhiming\_yu@sutd.edu.sg}
\affiliation{Research Laboratory for Quantum Materials, Singapore University of Technology and Design, Singapore 487372, Singapore}

\author{Hua Jiang}
\affiliation{College of Physics, Optoelectronics and Energy, Soochow University, Suzhou 215006, China}
\affiliation{Institute for Advanced Study, Soochow University, Suzhou 215006, China}

\author{Shengyuan A. Yang}
\affiliation{Research Laboratory for Quantum Materials, Singapore University of Technology and Design, Singapore 487372, Singapore}

\begin{abstract}
At a normal-metal/superconductor interface, an incident electron from the normal-metal (N) side can be normally reflected as an electron or Andreev reflected as a hole. We show that pronounced lateral shifts along the interface between the incident and the reflected quasiparticles can happen in both reflection processes, which are analogous to the Goos-H\"{a}nchen effect in optics. Two concrete model systems are considered. For the simplest model in which the N side is of the two-dimensional electron gas, we find that while the shift in Andreev reflection stays positive, the shift in normal reflection can be made either positive or negative, depending on the excitation energy. For the second model with the N side taken by graphene, the shift in Andreev reflection can also be made negative, and the shifts have rich behavior due to the additional sublattice pseudospin degree of freedom. We show that the shift strongly modifies the dispersion for the confined waveguide modes in an SNS structure. We also suggest a possible experimental setup for detecting the shift.
\end{abstract}

\maketitle

\section{Introduction}
The analogies between electronics and optics have inspired many breakthroughs in both fields. For example, the famous Datta-Das spin field effect transistor is inspired by the electro-optic light modulator~\cite{Datta1990}; and the concept of photonic crystal follows from the electronic band structure for crystalline solids~\cite{Joannopoulos2011}. Nowadays, with the rapid development of experimental techniques, the electron mean free path can even reach micron scale, giving rise to the flourishing field of electron optics~\cite{Spector1990,Molenkamp1990,Dragoman1999}, which may enable the exploration of more optical analogies in electronic systems.

There exists an interesting optical phenomenon: A light beam could acquire a longitudinal spatial shift \emph{within} the incident plane during a total reflection at an optical interface, known as the Goos-H\"{a}nchen (GH) effect~\cite{Goos1947}. The effect is a typical wave phenomenon and has been used as a powerful probe for interface properties in optics, acoustics, and atomic physics~\cite{Fornelevanescent2011}. It should be mentioned that a transverse shift perpendicular to the incident plane, known as the Imbert-Fedorov (IF) effect~\cite{Fedorov1955,Imbert1972}, may also occur when the light is circularly polarized. In this work, we focus on the electronic analogue of the GH effect, not the IF effect.

Several previous works have studied the GH-like shift for electrons scattered at scalar or magnetic potential barriers~\cite{Miller1972,Fradkin1974,Sinitsyn2005,Chen2008,Beenakker2009,Sharma2011,Wu2011,Chen2011}. The shift is closely related to the presence of evanescent modes, and typically diverges when approaching the total reflection angle. For massless Dirac electrons in graphene, it was found that the GH-like shift can be quite large due to the relatively small Fermi wavelength, and it has a strong dependence on the sublattice pseudospin degree of freedom~\cite{Beenakker2009}.

For all previously mentioned cases, the reflected quasiparticle retains the same identity as the incident one, i.e., an incident electron is reflected as an outgoing electron. Notably, there exists a special and intriguing type of fundamental reflection process---Andreev reflection~\cite{Andreev1964}, in which the quasiparticle identity is changed. The process happens at a normal-metal/superconductor (NS) interface. Besides the normal electron-to-electron reflection, an incident electron from the normal-metal (N) side can also be Andreev-reflected as a hole at the NS interface. Andreev reflection conserves energy and momentum but not charge---the two missing electrons are transferred into the superconductor as a Cooper pair. It is then natural and highly interesting to ask the following questions. (i) \emph{Is there a sizable GH-like shift also for Andreev reflections?} (ii) \emph{Is there any special feature in GH-like shift for normal reflections at an NS interface?}

In this paper, we try to address the above two questions. This work is also motivated by our recent discovery of sizable IF-like transverse shift in Andreev reflection~\cite{Liu2017,Yu2017}. There, the presence of GH-like shift was noticed but not investigated in detail~\cite{Liu2017}. We also notice a previous theoretical work which provided a negative answer for question (i)~\cite{Lee2013}. Here, we show that Ref.~\cite{Lee2013}'s conclusion on the absence of GH-like shift applies for a particular limit but not the general case. By studying two concrete model systems, we show that pronounced GH-like shifts can happen in both Andreev and normal reflection processes at an NS interface. In the first model, the N side is of the two-dimensional electron gas (2DEG). We find that while the shift in Andreev reflection stays positive, the shift in normal
reflection can be made either positive or negative by tuning the excitation energy. The second model is for a graphene NS junction, where we find that the shift in Andreev reflection can also be made negative, and the shifts have rich behavior due to the additional sublattice pseudospin degree of freedom. We show that the shift modifies the dispersion for the confined waveguide modes in an SNS structure.  A possible experimental setup for detecting the anomalous shift is suggested.

\section{General formulation}
\begin{figure}[t]
\includegraphics[width=8cm]{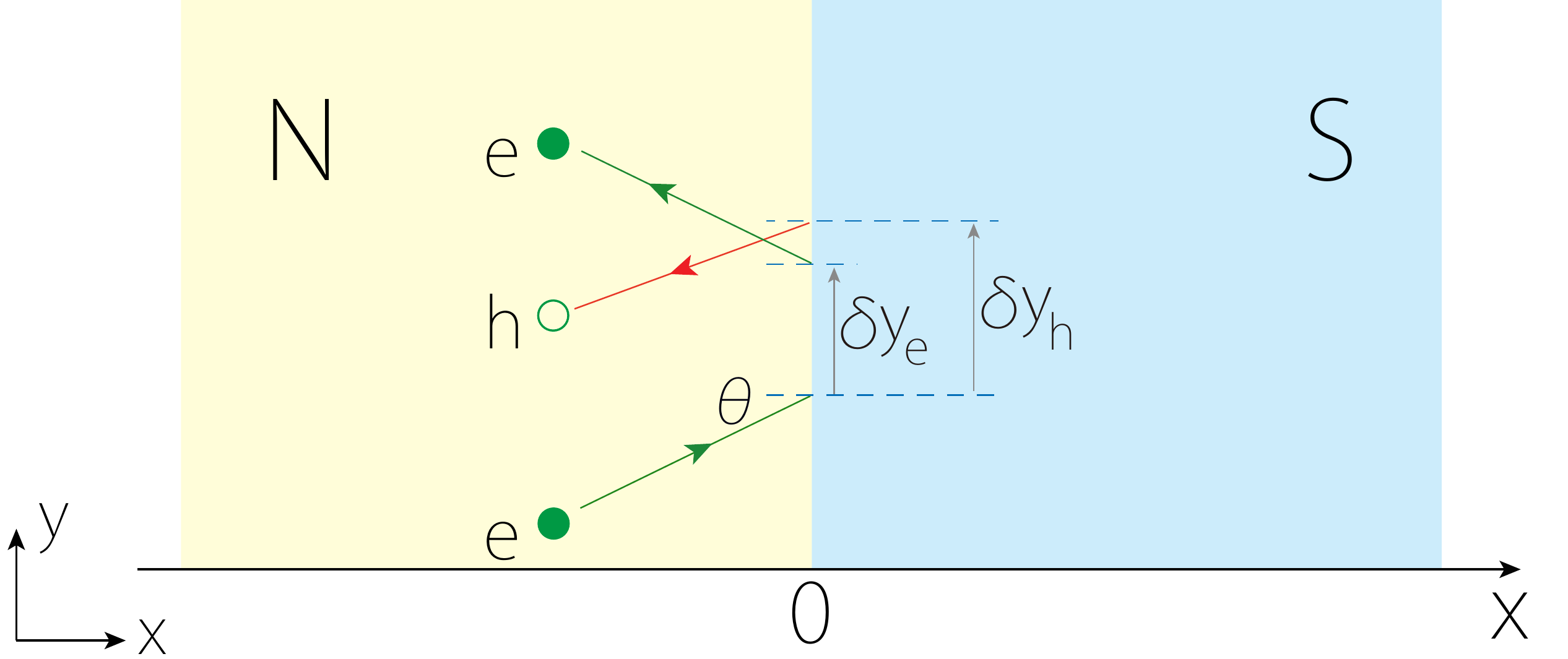}
\caption{Schematic figure showing the GH-like shifts in normal reflection ($\delta {y}_e$) and in Andreev reflection ($\delta {y_h}$) for an incident electron beam reflected from an NS interface. The solid and open circles indicate the electron and the hole quasiparticles, respectively. }
\label{Fig_sch}
\end{figure}

Consider a clean and flat NS interface located at $x=0$, as illustrated in Fig.~\ref{Fig_sch}. Since the GH-like shift is within the incident plane (here the $x$-$y$ plane), we may just consider a two-dimensional setup for the investigation. This is unlike the study of IF-like shift, which necessarily requires a three-dimensional setup~\cite{Yang2015}.  In our model, the $x<0$ region is for the normal metal, whereas $x>0$ is for the superconductor. We assume that the system is uniform along the $y$ direction.
The scattering at the NS interface for the quasiparticle is described by the microscopic Bogoliubov-de Gennes (BdG) equation~\cite{Gennes1966,Blonder1982}:
\begin{equation}
  \left[\begin{array}{cc}
         H_0+U(\bm r)-E_F & \Delta(\bm r) \\
         \Delta^*(\bm r) & E_F-\mathcal T^{-1}H_0\mathcal T-U(\bm r)
       \end{array}
  \right]\psi=\varepsilon\psi. \label{BdG}
\end{equation}
Here, we assume that the S side is of the conventional $s$-wave pairing and the real-spin labels are suppressed, $H_0$ is the Hamiltonian in the normal (non-superconducting) state,  $U(\bm r)=-U_0\Theta(x)$ denotes a potential energy offset (which may be tuned by doping or electric gating) between the two regions with $\Theta(x)$ the Heaviside step function, $\Delta(\bm r)=\Delta_0\Theta(x)$ represents the pair potential on the S side, $\varepsilon$ is the quasiparticle excitation energy measured from the Fermi level, and  ${\cal T}$ is the time reversal operator. Here, for a single NS interface, the superconducting phase is inessential and can always be gauged away.
The wavefunction $\psi=(u,v)^T$ is a mutli-component spinor with  $u$ ($v$)  standing for electron (hole) state.
The mean-field requirement of superconductivity is that $E_F+U_0\gg\Delta_0$ on the S side, i.e., the Fermi
wavelength in the S region should be much smaller than the coherence length. Meanwhile, the Fermi wavelength on the N side is not constrained to be small, e.g., $E_F$ can be comparable to $\Delta_0$ provided $U_0$ is large.

An incoming electron from the N side is scattered at the interface. The scattering properties are captured by the scattering amplitudes, which can be obtained by solving the scattering states of the BdG equation~\cite{Blonder1982}. The typical scattering state takes the following form:
\begin{eqnarray}
\psi(\boldsymbol{r}) & = & \begin{cases}
\psi^{e+}+r_e\psi^{e-}+r_{h}\psi^{h-}, & x<0,\\
t_{+}\psi_{+}^\text{S}+t_{-}\psi_{-}^\text{S}, & x>0,
\end{cases}
\end{eqnarray}
where $\psi^{e+}$ is the incident electron state, $\psi^{e(h)-}$ is the reflected electron (hole) state on the N side, $\psi_{\pm}^\text{S}$ are the transmitted quasiparticle states on the S side, $r_{e(h)}$ is the reflection amplitude for the normal (Andreev) reflection, and $t_\pm$ are the amplitudes for transmission. Note that the energy ($\varepsilon$) and the momentum along the interface ($k_y$) are conserved in the scattering process.

The anomalous positional shift in scattering is defined for a quasiparticle beam. Following the standard quantum scattering approach~\cite{Beenakker2009,Jiang2015,Liu2017}, this is modeled by a quasiparticle wave-packet, which can be written as
\begin{eqnarray}
	\Psi^{e+}(\bm r^c,\bm k^c)=\int d \bm k w(\bm k-\bm k^c)\psi^{e+}(\bm k)
\end{eqnarray}
for the incident electron. Here the profile $w$ ensures the wave-packet is peaked at the centroid ($\bm r^c$, $\bm k^c$) in phase space. In calculations, one usually chooses $w$ to have a Gaussian form: $w(\bm k-\bm k^c)=\prod_iw_i(k_i-k_i^c)$, where $w_i(k_i-k_i^c)=(\sqrt{2\pi}W_i)^{-1}\mathrm{exp}[-(k_i-k_i^c)^2/(2W_i^2)]$, with a width $W_i$ for the $i$-th component. However, it should be noted that the obtained shift does not depend on the specific form of the profile. Each partial wave $\psi^{e+}$ is scattered at the interface according to the scattering amplitudes. The reflected electron (hole) wave-packet is then given by $\Psi^{e(h)-}(\bm r)=\int d \bm k w(\bm k-\bm k^c)r_{e(h)}\psi^{e(h)-}(\bm k)$.

The GH-like shift can be obtained by comparing the center positions for the incident and the reflected beams at the interface.
For example, for the 2DEG/superconductor model to be discussed in Sec.~III, by expanding the phase of the amplitude $r_{e(h)}$ to the first order around $k^c_y$, the $k$-integral of the wave-packet gives that
$
\Psi^{e(h)-} \propto   e^{-W_{y}^{2} [r_y+\frac{\partial}{\partial k_y}\arg (r_{e(h)})|_{k^c_y}]^2/2}.
$
Compared with the incident electron
$
\Psi^{e+} \propto  e^{-W_{y}^{2} r_y^2/2},
$
one finds that the reflected electron (hole) has a relative spatial shift along the $y$ direction of
\begin{eqnarray}
\delta {y}_{e(h)}= -\frac{\partial}{\partial k_y}\arg (r_{e(h)})\Big|_{k^c_y}.\label{GasGH}
\end{eqnarray}
The shift for graphene/superconductor model in Sec.~IV can be obtained in a similar way, except that the shift needs to be averaged over the components of the electron (hole) spinor states~\cite{Beenakker2009}, due to the additional pseudospin degree of freedom.
A straightforward calculation shows that the final expression of the GH-like shift for the graphene/superconductor model also takes form in Eq.~(\ref{GasGH}).

There are two remarks before proceeding. First, in this quantum scattering approach, it is clear that the shift arises from the interference between scattered partial waves that undergo different $k$-dependent phase shifts [see Eq.~(\ref{GasGH})]. Second, for certain simple cases, the shift may also be derived from the semiclassical equations of motion for the wave-packet~\cite{Yang2015}. However, in that semiclassical approach, the variation of the scattering potential is required to be slow compared with the Fermi wavelength. In comparison, the quantum scattering approach here does not have such constraint. It applies to the cases with small Fermi wavelengths and sharp interfaces as well~\cite{Jiang2015,Liu2017}.

\section{Model I: 2DEG/superconductor Junction}

\begin{figure}[t]
\includegraphics[width=8.4cm]{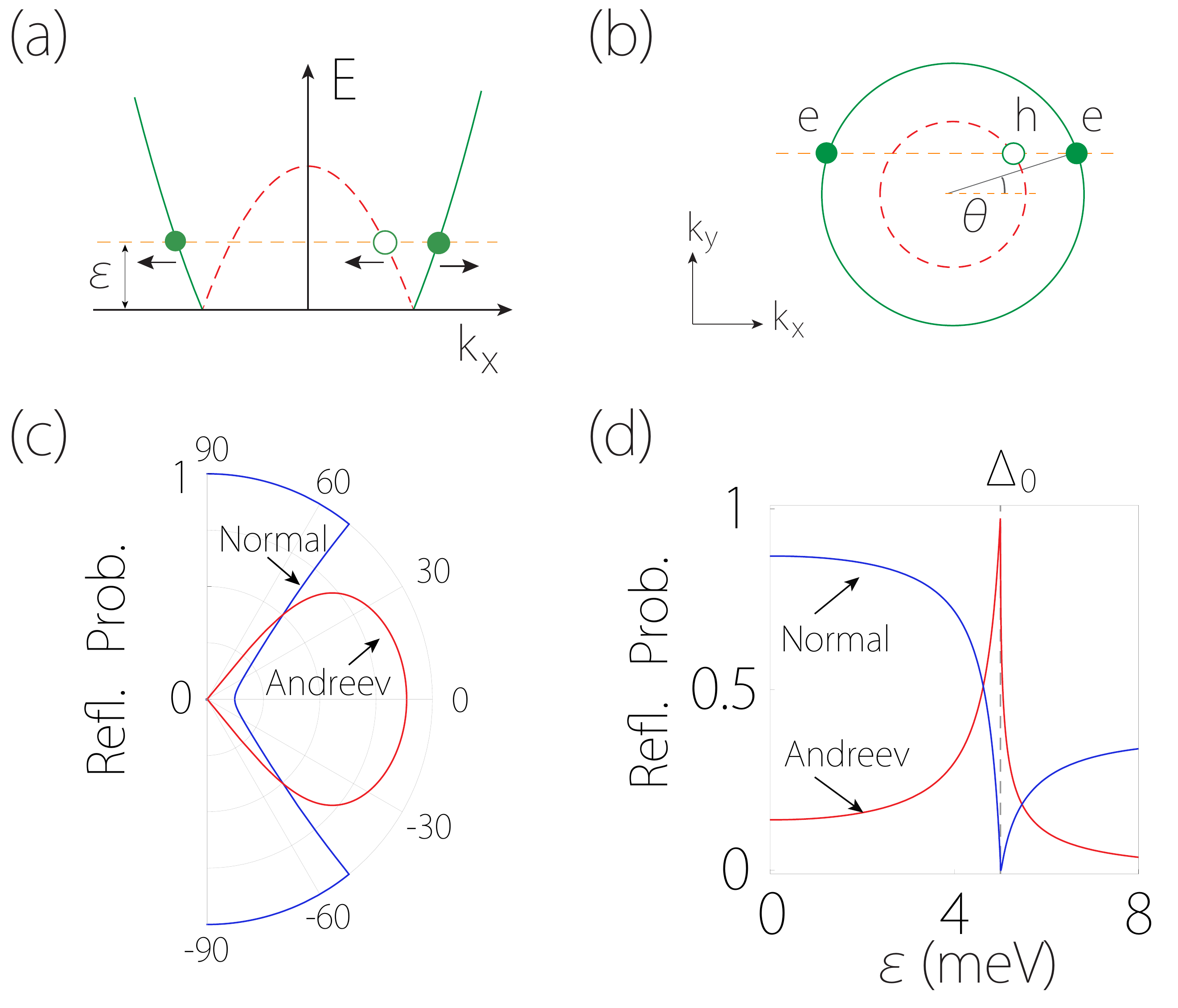}
\caption{(a,b) Schematic figure showing (a) the BdG dispersion and (b) the equi-energy contours in the N region for the 2DEG/superconductor model. The solid (open) sphere denotes the electron (hole) state and the arrows indicate the propagation directions. The green solid (red dashed) curve indicates the electron (hole) band. (c,d) Probabilities for normal reflection (bule curve) and Andreev reflection (red curve) as functions of (c) the incident angle $\theta$ and (d) the excitation energy $\varepsilon$. In (c) and (d), we set $U_0=500$ meV, $E_F=20$ meV, $\Delta_0=5$ meV, $m=0.05\  m_e$. We take the excitation energy $\varepsilon=4.95$ meV in (c), and the incident angle $\theta=\pi/10$ in (d).  }
\label{Fig_EG}
\end{figure}

The first model that we consider is of a 2DEG interfaced with a superconductor. Here, the Hamiltonian $H_0$ is given by (set $\hbar=1$)
\begin{equation}
  H_0=\frac{1}{2m} k^2,
\end{equation}
with $m$ the effective mass of the electron. The BdG spectrum for the N side is schematically shown in Fig.~\ref{Fig_EG}(a). One observes that for low excitation energies with $\varepsilon<E_F$, there are two equi-energy contours belonging to the electron and the hole bands respectively [see Fig.~\ref{Fig_EG}(b)], and hence an incident electron can be normal-reflected as an electron or Andreev-reflected as a hole, as illustrated in Fig.~\ref{Fig_EG}(a). However, from Fig.~\ref{Fig_EG}(b), due to the conservation of $k_y$, for incident angle $\theta$ greater than a critical value given by
\begin{equation}
\theta_c=\arcsin\sqrt{\epsilon_-/\epsilon_+},
\end{equation}
the Andreev reflection is no longer allowed, where $\epsilon_\pm=E_F\pm \varepsilon$. When $E_F\gg \varepsilon$, $\theta_c$ approaches $\pi/2$, and Andreev reflection is allowed for all incident angles. For smaller $E_F$ (but still $E_F>\varepsilon$), Andreev reflection is allowed only for $|\theta|<\theta_c$. Finally, for the case with $\varepsilon>E_F$, the equi-energy contour in the hole band disappears (for which we may set $\theta_c=0$), and only normal reflection is allowed.

The electron and hole basis states in the N region are given by
\begin{eqnarray}
\psi^{e\pm}&=&\sqrt{\frac{m}{k_e}}\left(\begin{array}{c}
                       1 \\
                       0
                     \end{array}
  \right)e^{\pm ik_e x+ik_yy},\\
\psi^{h-}&=&\sqrt{\frac{m}{k_h}}\left(\begin{array}{c}
                       0 \\
                       1
                     \end{array}
  \right)e^{ik_h x+ik_yy},
\end{eqnarray}
where $k_e=(2m\epsilon_+ -k_y^2)^{1/2}$ is the $x$-component of the electron wave-vector (here we drop the subscript $x$ for simple notations), $k_h=(2m\epsilon_- -k_y^2)^{1/2}$ for $|\theta|<\theta_c$, whereas $k_h=-i(k_y^2-2m\epsilon_-)^{1/2}$ for $|\theta|>\theta_c$.
The normalization factors here are chosen to ensure that the propagating states in the N region carry the same particle current.

Meanwhile, the basis states in the S region read
\begin{equation}
  \psi^S_\pm=\left(\begin{array}{c}
                     e^{\pm i\beta} \\
                     1
                   \end{array}
  \right)e^{\pm ik_{S} x-\kappa x+ik_yy},
\end{equation}
where  $k_S=[2m(E_F+U_0)-k_y^2]^{1/2}$, $\kappa=m\Delta_0\sin\beta/k_S$, and $\beta=\arccos(\varepsilon/\Delta_0)$ for $\varepsilon<\Delta_0$, whereas $\beta=-i\cosh^{-1}(\varepsilon/\Delta_0)$ for $\varepsilon>\Delta_0$.
We have used the assumed condition that $(E_F+U_0)\gg\Delta_0, \varepsilon$.

With the boundary conditions at the interface ($x=0$)
\begin{eqnarray}
\psi\big|_{x=0^+}&=&\psi\big|_{x=0^-},\\
\partial_x\psi\big|_{x=0^+}&=&\partial_x\psi\big|_{x=0^-},
\end{eqnarray}
one can solve for the scattering amplitudes. Particularly, we find that the two reflection amplitudes take the following form:
\begin{eqnarray}
  r_e&=&X^{-1}\Big[k_S(k_e-k_h)\cos\beta + \kappa(k_e+k_h)\sin\beta \nonumber\\
  &&+i(k_ek_h-k_S^2-\kappa^2)\sin\beta\Big]\label{Ncoe_EG}
\end{eqnarray}
and
\begin{equation}
  r_h=2X^{-1}k_S\sqrt{k_e k_h}\label{Acoe_EG}
\end{equation}
where $X=k_S(k_e+k_h)\cos\beta+\kappa(k_e-k_h)\sin\beta+i(k_ek_h+k_S^2+\kappa^2)\sin\beta$.
One checks that when $\varepsilon<\Delta_0$ and {$|\theta|<\theta_c$,} we have $|r_e|^2+|r_h|^2=1$ as required by the conservation of quasiparticle current, because there is no quasiparticle transmission into the S side.

The GH-like shifts can then be directly calculated using Eq.~(\ref{GasGH}), but the results are quite complicated. To gain better understanding, let's consider two different regimes. First, when the N side is heavily doped such that $E_F\gg U_0, \Delta_0, \varepsilon$, the reflection amplitudes reduce to
\begin{eqnarray}
	r_e=0, \ \ \ r_h=e^{-i\beta}.
\end{eqnarray}
In this regime, the Andreev reflection dominates the scattering. Importantly, the phase shift $\arg(r_{e/h})$ is a $k$-independent value, so that the GH-like shifts for both normal and Andreev reflections vanish: $\delta y_{e/h}=0$. This recovers the result obtained in Ref.~\cite{Lee2013} which assumed this regime.

However, outside of the above regime, when $E_F$ is not large, the shifts would generally be nonzero. Let's consider the more interesting regime when the N side is lightly doped (compared with the S side) such that $U_0 \gg E_F, \varepsilon, \Delta_0$. Then the reflection amplitudes are reduced to
\begin{eqnarray}
r_e&=&{X^{\prime}}^{-1}\left[ (k_e-k_h)\cos\beta-ik_S \sin\beta \right],\label{Ncoe_LEG}\\
r_h&=&2{X^{\prime}}^{-1}\sqrt{k_e k_h},\label{Acoe_LEG}
\end{eqnarray}
with
$
X^{\prime}=(k_e+k_h)\cos\beta+ik_S\sin\beta
$. In Fig.~\ref{Fig_EG}(c), (d), we plot the reflection probabilities for the two processes as functions of the incident angle and the excitation energy. One observes that due to the Fermi surface mismatch between the two sides, the Andreev reflection probability is decreased, and there is a competition between the two processes. From Fig.~\ref{Fig_EG}(d), one also observes that the probability for Andreev reflection is typically large when $\varepsilon\lesssim\Delta_0$~\cite{Blonder1982}.

\begin{figure}[t]
\includegraphics[width=8.4cm]{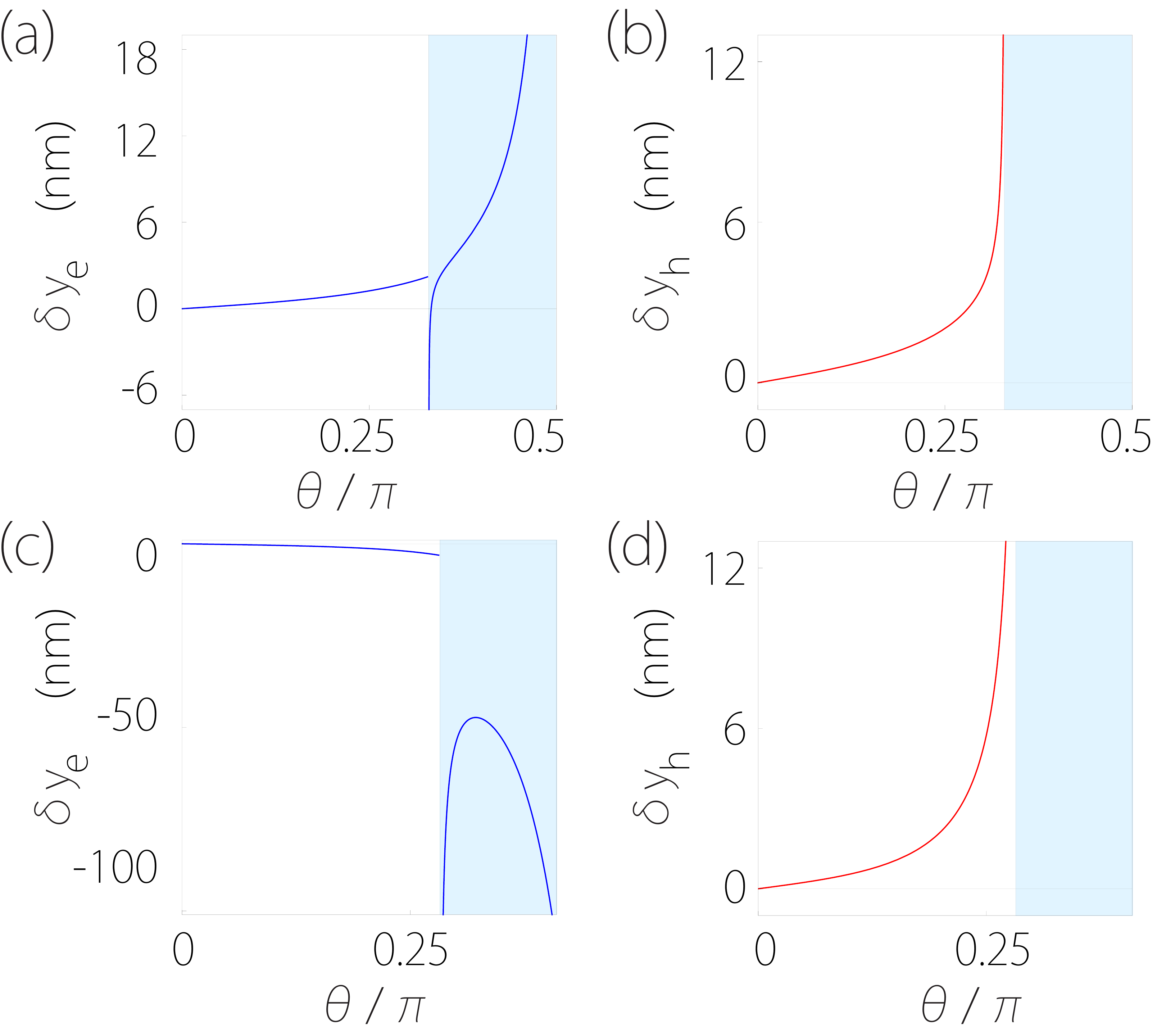}
\caption{GH-like shifts (a,c) in the normal reflection ($\delta y_e$), and (b,d) in the Andreev reflection ($\delta y_h$) versus the incident angle $\theta$ for the 2DEG/superconductor model. (a,b) are with $\varepsilon=3$ meV, and (c,d) are with $\varepsilon=4.98$ meV. The shadowed region in each figure denotes the range with $|\theta|>\theta_c$, where Andreev reflection is not allowed.  Here, we take  $E_F=20$ meV, $U_0=500$ meV, $\Delta_0=5$ meV, and $m=0.05\ m_e$.}
\label{Fig_EGsft}
\end{figure}

In the second regime, we are most interested in the case when $\varepsilon<\Delta_0$, because the reflections are dominating the interface scattering. Substituting Eqs.~(\ref{Ncoe_LEG}) and (\ref{Acoe_LEG}) into Eq.~(\ref{GasGH}), we find that for $|\theta|<\theta_c$, the shifts are given by
\begin{eqnarray}
\delta {y}_e&=&
        \frac
        {2\tan\theta(k_h^2-k_e^2+k_S^2\tan^2\beta)k_S\tan\beta}
        {(k_h^2+k_e^2+k_S^2\tan^2\beta)^2-4k_e^2 k_h^2},\label{EGsft1}
\end{eqnarray}
and
\begin{eqnarray}
\delta {y_h}&=&
\frac{\tan\theta}{k_h} \cdot
\frac{(k_e+k_h)k_S\tan\beta}{(k_e+k_h)^2+k_S^2\tan^2\beta};\label{EGsft2}
\end{eqnarray}
whereas for $|\theta|>\theta_c$ (including the case with $\varepsilon>E_F$), the Andreev reflection is not allowed, and
\begin{eqnarray}
\delta {y}_e &=&
\frac{2\tan\theta}{|k_h|}\cdot
\frac
{|k_h| k_S\tan\beta-(k_e^2+|k_h|^2)}
{k_e^2+(k_S \tan \beta- |k_h|)^2}. \label{EGsft3}
\end{eqnarray}

We plot the typical behavior of the shifts in Fig.~\ref{Fig_EGsft}. From the results, we observe the following features. First, the shifts are odd functions of the incident angle $\theta$, as it should be, due to the mirror symmetry of the system (mirror plane perpendicular to $y$). Consequently, the shifts should vanish for normal incidence ($\theta=0$). Second, the shift in Andreev reflection is always in the forward direction for this model, i.e., for $\theta>0$, we have $\delta y_h>0$ according to Eq.~(\ref{EGsft2}). This shift can be large when $|\theta|\lesssim \theta_c$. Third, interestingly, the shift in normal reflection can be either positive or negative. Particularly, its magnitude is very large in the angular range $|\theta|>\theta_c$ where only normal reflection is allowed, and pronounced backward shift is observed when the excitation energy is close to the superconducting gap ($\varepsilon\lesssim\Delta_0$) [see Fig.~\ref{Fig_EGsft}(c)]. Finally, there is a discontinuity in $\delta y_e$ at the critical angle $\theta=\theta_c$, above which $\delta y_e$ is divergingly large. This is reminiscent of the diverging GH shift near the total reflection angle~\cite{Miller1972}. Physically, this is because across the critical angle, the original propagating mode for the hole becomes evanescent, which strongly affects the phase shifts also in the normal reflection channel.

To better understand the pronounced backward shift in normal reflection discussed above (the third feature), we note that for $|\theta|>\theta_c$,
$r_e=X'^*/X'$, with $X'=k_e\cos\beta+i(k_S\sin\beta-|k_h|\cos\beta)$. The key observation is that the amplitude $k_h$ now turns imaginary, and contributes to the imaginary part of $X'$. Hence, the phase shift in normal reflection becomes
\begin{equation}
\arg(r_e)=-2\arctan\left(\frac{k_S\sin\beta-|k_h|\cos\beta}{k_e\cos\beta}\right).
\end{equation}
When $\varepsilon\lesssim\Delta_0$, $\beta$ is close to zero, so $ \arg(r_e)\approx 2\arctan(|k_h|/k_e)$, from which one directly finds that
\begin{eqnarray}\label{yesim}
\delta y_e&\approx&    -\frac{2\tan\theta}{|k_h| },
\end{eqnarray}
clearly showing that the shift is in the backward direction and it diverges when $\theta$ approaches $\theta_c$ from above since $|k_h|$ approaches zero. It is interesting to see that for $|\theta|>\theta_c$, although Andreev reflection itself is prohibited, the amplitude $k_h$ for the hole (evanescent) mode turns out to strongly affect the shift in normal reflection, as reflected in Eq.~(\ref{yesim}).

\section{Model II: Graphene/superconductor Junction}

\begin{figure}[t]
\includegraphics[width=8.4cm]{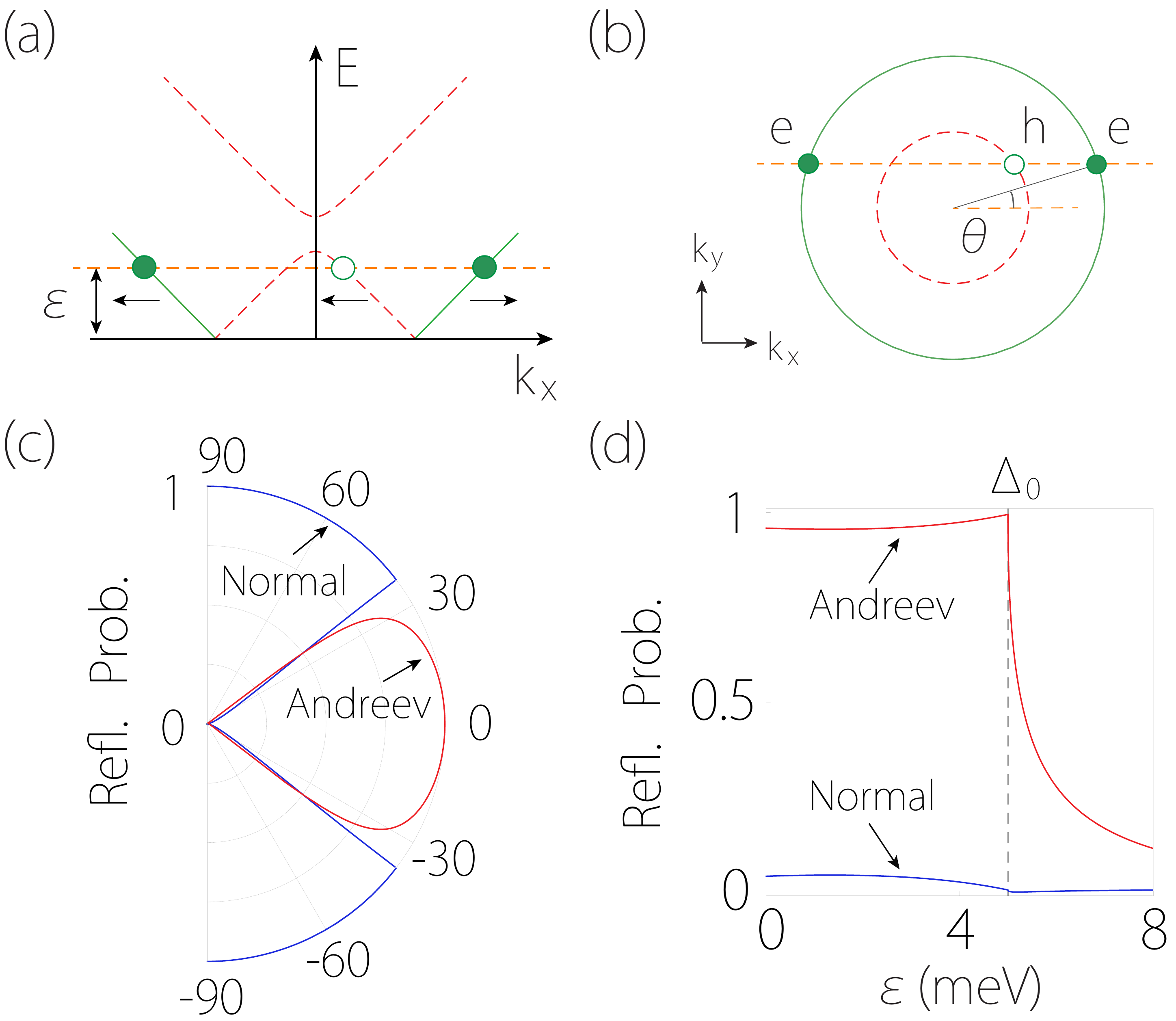}
\caption{(a,b) Schematic figure showing (a) the BdG dispersion and (b) the equi-energy contours in the N region for the graphene/superconductor model. In (a), there appears a gap in the spectrum, because it is plotted for a finite $k_y$. In (a,b), the green solid (red dashed) curve indicates the electron (hole) band, and we plot the case with $\varepsilon<E_F$. (c,d) Probabilities for the normal reflection (blue curve) and the Andreev reflection  (red curve) as functions of (c) the incident angle $\theta$, and (d) the excitation energy $\varepsilon$. Here, we set $U_0=500$ meV, $\Delta_0=5$ meV, $E_F=20$ meV, $v_F=1\times 10^6$ m/s. We take $\varepsilon=4.9$ meV in (c), and $\theta=\pi/10$ in (d).  }
\label{Fig_gra}
\end{figure}

In the second model, we consider an NS junction based on graphene~\cite{Novoselov2004}. Let us first recall some basic facts about graphene. It is a single sheet of carbon atoms, featuring Dirac-cone like dispersion in its low-energy band structure~\cite{Neto2009}. There are two Dirac cones (valleys) located at the two corner points $\pm K$ of the hexagonal Brillouin zone, which are connected by the time reversal symmetry. The low-energy Hamiltonian for graphene takes the form of~\cite{Neto2009}
\begin{eqnarray}\label{HG}
	&H_0^{\tau}(\bm k)=v_F(\tau k_x\sigma_x+k_y\sigma_y),
\end{eqnarray}
where $\tau=\pm$ denotes the two valleys, $v_F$ is the Fermi velocity, the wave-vector is measured from the valley center, and $\sigma$'s are the Pauli matrices acting on $A/B$ sublattice degree of freedom.

In modeling the junction, we assume that the S region is also described by the Hamiltonian (\ref{HG}) but with a nonzero pair potential (and a potential energy offset). Physically, this may be realized by covering the graphene in the S region with a superconducting electrode, which induces a finite $\Delta$ by proximity effect. The potential energy offset $U$ may be adjusted by gate voltage or by doping. This model has been used by Beenakker~\cite{Beenakker2006} in discussing the special specular Andreev reflection in graphene.

Now we substitute $H_0$ in Eq.~(\ref{HG}) into the BdG equation in Eq.~(\ref{BdG}). We notice that
$\mathcal TH_0^{\tau}(\bm k)\mathcal T^{-1}=H_0^{-\tau}(-\bm k)$, indicating that an incident electron in one valley is coupled to the hole in the other valley through the superconducting pair potential in the S region. Neglecting the intervalley scattering (due to the large separation between the two valleys in $k$-space), we can write the BdG equation into two decoupled sets~\cite{Beenakker2006}, and the wavefunction $\psi$ in the equation then takes a four-component form. For the case with electron components in the $K$ valley, we have $\psi =(u_{A+}, u_{B+},v_{A-},v_{B-})^T$, where the subscript $A/B$ stands for the two sublattices and $\pm$ denotes the two valleys.

The BdG spectrum and the equi-energy contours of the N region (for a single valley) are schematically shown in Fig.~\ref{Fig_gra}(a) and \ref{Fig_gra}(b). Due to the Dirac-cone spectrum of graphene, there always exist propagating hole states for Andreev reflection at all energies. And similar to the 2DEG/superconductor model, here we also have a critical angle $\theta_c$ for Andreev reflection. From Fig.~\ref{Fig_gra}(b), one easily finds that
\begin{eqnarray}
\theta_c=\arcsin(\epsilon_-/\epsilon_+),
\end{eqnarray}
where $\epsilon_{\pm}=E_F\pm \varepsilon$ has the same definition as in the previous section, and $E_F$ here is measured from the Dirac point.
For incident angle $|\theta|>\theta_c$, the Andreev reflection is prohibited.

Consider an incident electron from the $K$ valley (the analysis and result below also apply for the case of $K'$ valley). The basis states for the incident and the reflected electrons in the N region are
\begin{eqnarray}
	\psi^{e\pm}&=&\frac{1}{\sqrt{\cos\theta}}\left[\begin{array}{c}
	e^{\mp i\theta/2}\\
	\pm e^{\pm i\theta/2}\\
	0\\
	0
	\end{array}\right]e^{\pm ik_e x+ik_yy}\label{Ewave},
\end{eqnarray}
and the basis state for the reflected hole is
\begin{eqnarray}
\psi^{h-}&=&\frac{1}{\sqrt{\cos\theta_h}}\left[\begin{array}{c}
0\\
0\\
e^{-i\theta_h/2}\\
e^{i\theta_h/2}\\
\end{array}\right]e^{ik_h x+ik_yy},\label{Hwave}
\end{eqnarray}
where $k_e=(\epsilon_{+}^2/v_F^2-k_y^2)^{1/2}$, $\theta_{h}=\arcsin(v_F k_y/\epsilon_{-})$, $k_{h}=\mathrm{sgn}(\epsilon_{-})(\epsilon_{-}^2/v_F^{2}-k_y^2)^{1/2}$ for $|\theta|<\theta_c$, and $k_h=-i (k_y^2-\epsilon_{-}^2/v_F^{2})^{1/2}$ for $|\theta|>\theta_c$. The normalization factors
are added to ensure that the propagating states in the N region carry the same particle current.

The basis states for the S region are given by
\begin{eqnarray}
	\psi^S_\pm=\left[\begin{array}{c}
	e^{\pm i\beta}\\
	\pm e^{\pm i(\beta + \beta^{\prime})}\\
	1\\
	\pm e^{\pm i \beta^{\prime}}
	\end{array}\right]e^{\pm ik_S x-\kappa x+ik_yy},\label{Swave}
\end{eqnarray}
where $k_S=[(E_F+U_0)^2/v_F^2- k_y^2]^{1/2}$, $\kappa=(E_F+U_0)\Delta_0\sin\beta/(v_F^2 k_S)$, $\beta^{\prime}=\arctan (k_y/k_S)$, $\beta=\arccos(\varepsilon/\Delta_0)$ for $\varepsilon<\Delta_0$, and  $\beta=-i\cosh^{-1}(\varepsilon/\Delta_0)$ for $\varepsilon>\Delta_0$. Here, again, the condition  $(E_F+U_0)\gg\Delta_0, \varepsilon$  is assumed.

The boundary condition here is given by the continuity of the wavefunction at the interface:
\begin{eqnarray}
	\psi(x=0^+)=\psi(x=0^-),
\end{eqnarray}
from which the scattering  amplitudes can be solved.

Like for the previous model, we first consider the regime when the N region is heavily doped, with $E_F\gg U_0, \Delta_0, \varepsilon$. In this case, we find that
\begin{eqnarray}
	r_e=0, \ \ \ r_h=e^{-i\beta},
\end{eqnarray}
same as for the 2DEG/superconductor model. Since the phase shifts are $k$-independent, the GH-like shifts vanish in this regime.

In the following, we  focus on the regime where the N region is lightly doped, such that $U_0\gg E_F, \Delta_0, \varepsilon$.
A straightforward calculation yields the following results for the reflection amplitudes:
\begin{eqnarray}
	r_e&=&-Y^{-1}(\sin\beta\sin \theta_{+} -i\cos \beta \sin \theta_{-}), \label{Gncoe} \\
	r_h&=& Y^{-1}\sqrt{\cos\theta \cos\theta_h}\label{Gacoe},
\end{eqnarray}
where  $\theta_{\pm}=(\theta \pm \theta_h)/2$, and
$
Y=\cos\beta\cos \theta_{+}+i\sin\beta\cos \theta_{-}.
$
The probabilities for the two reflection processes are plotted in Fig.~\ref{Fig_gra}(c) and \ref{Fig_gra}(d). One observes that close to normal incidence, the normal reflection is suppressed, because the reflected state has a reversed sublattice pseudospin. So the probability for Andreev reflection is quite large in this case. As observed from Fig.~\ref{Fig_gra}(d), the Andreev reflection also dominates when $\varepsilon \lesssim\Delta_0$.
In addition, one verifies that the results in Eq.~(\ref{Gncoe}) and (\ref{Gacoe}) satisfy the quasiparticle conservation relation
$|r_e|^2+|r_h|^2=1$, when $\varepsilon<\Delta_0$ and {$|\theta|<\theta_c$}.

\begin{figure}[t]
\includegraphics[width=8.6cm]{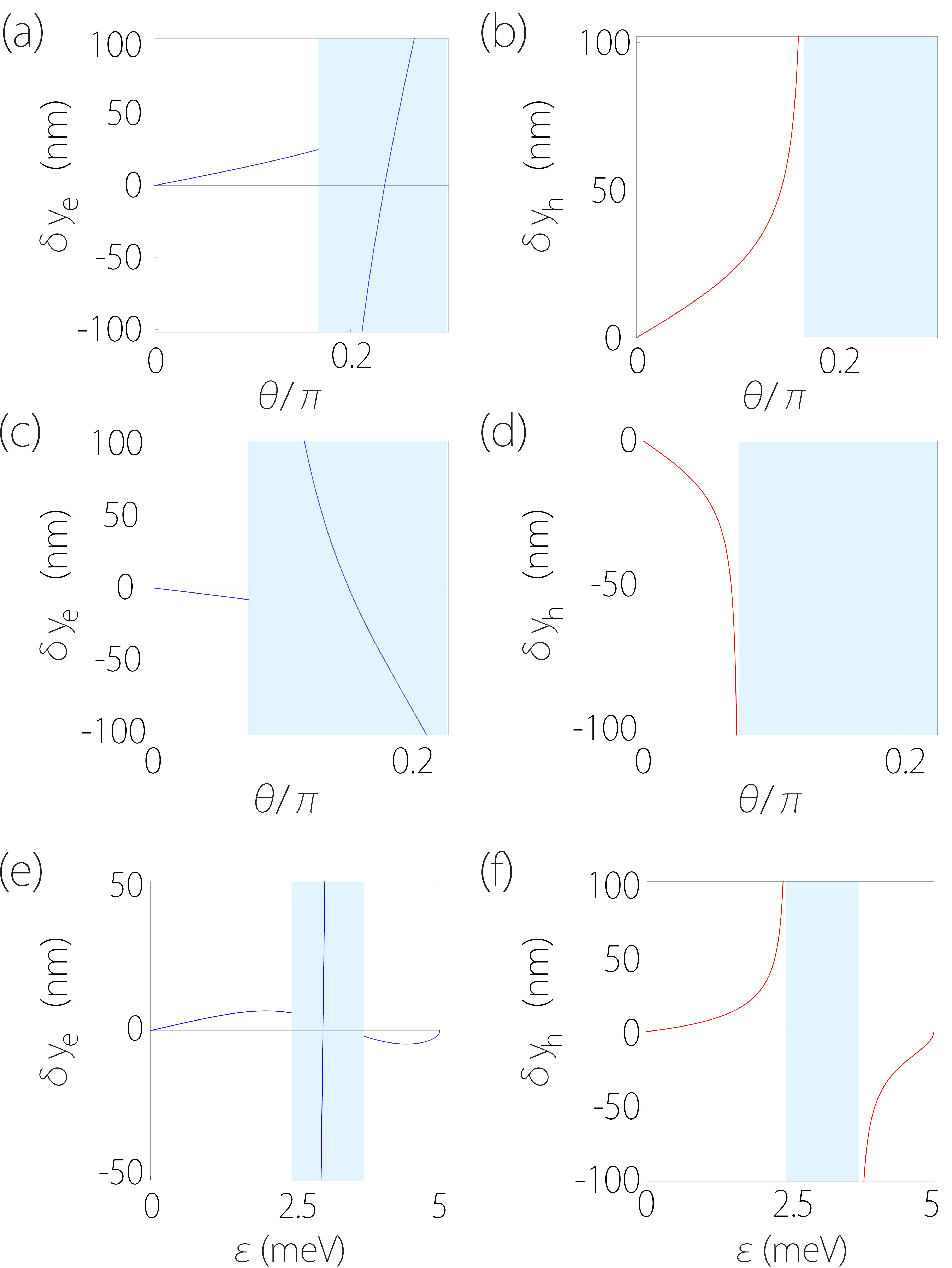}
\caption{GH-like shifts (a,c) in the normal reflection ($\delta {y_e}$), and (b,d) in the Andreev reflection ($\delta {y}_h$), as functions of the incident angle for the graphene/superconductor model.
(a,b) are for the case with $\varepsilon<E_F$ (with $E_F=3$ meV and $\varepsilon=1$ meV).
(c,d) are for the case with $\varepsilon>E_F$ (with $E_F=3$ meV, $\varepsilon=4.5$ meV).
(e,f)  GH-like shifts versus the excitation energy $\varepsilon$.
The shaded region in each figure denotes the region of $|\theta|>\theta_c$ where the Andreev reflection is not allowed. In these figures, we set $U_0=500$ meV, $\Delta_0=5$ meV and $v_F=1\times 10^{6}$ m/s. In (e,f), we take $E_F=3$ meV and $\theta=\pi/30$.}
\label{Fig_grasft}
\end{figure}

The GH-like shifts are obtained by combining Eq.~(\ref{Gncoe}) and (\ref{Gacoe}) with  Eq.~(\ref{GasGH}). Here, again, we are most interested in the case with $\varepsilon<\Delta_0$, for which the reflections are dominating. In this case, for $|\theta|<\theta_c$, we find
\begin{eqnarray}
\delta{y}_e & = &
\frac
{v_{F}\tan\theta\tan\beta (E_F\tan^2\beta-\varepsilon)}
{\epsilon_{+}^{2}\cos^2 \theta\tan^{2}\beta +(E_F\tan^2\beta-\varepsilon)^{2}}, \label{grasft1}
\end{eqnarray}
and
\begin{eqnarray}
	\delta{y_h}=
\frac{\tan\theta (\cos \theta+\cos \theta_h)}
{2k_h(\cot\beta\cos^2\theta_{+}+\tan\beta\sin^2\theta_{-})}. \label{grasft2}
\end{eqnarray}
For $|\theta|>\theta_c$, Andreev reflection is not allowed, and
\begin{eqnarray}
\delta{y}_e & = & \frac{2}{|k_{h}| \sin (2\theta) Z}
 \Big[\epsilon_{+}\epsilon_{-}\cos^{2}\theta-\epsilon_{+}^{2}\sin^{2}\theta\sin^{2}\beta  \nonumber\\
&&+(\epsilon_{-}\sin\beta-v_{F}|k_{h}|\cos\beta)^{2} \Big],\label{grasft3}
\end{eqnarray}
where $Z=\epsilon_{+}[\epsilon_{-}\cos(2\beta)-\epsilon_{+}+v_{F} |k_{h}|]$.

In Fig.~\ref{Fig_grasft}, we plot the results for these shifts. One observes that with reasonable parameter values, the magnitudes of the shifts here are quite large, even up to hundreds of nm. Several features are similar to that of the 2DEG/superconductor model. For example, these shifts are odd functions of the incident angle $\theta$; the shift in normal reflection ($\delta y_e$) has discontinuity at $\theta_c$, above which its value is divergingly large.

\begin{figure}[t]
\includegraphics[width=8.8cm]{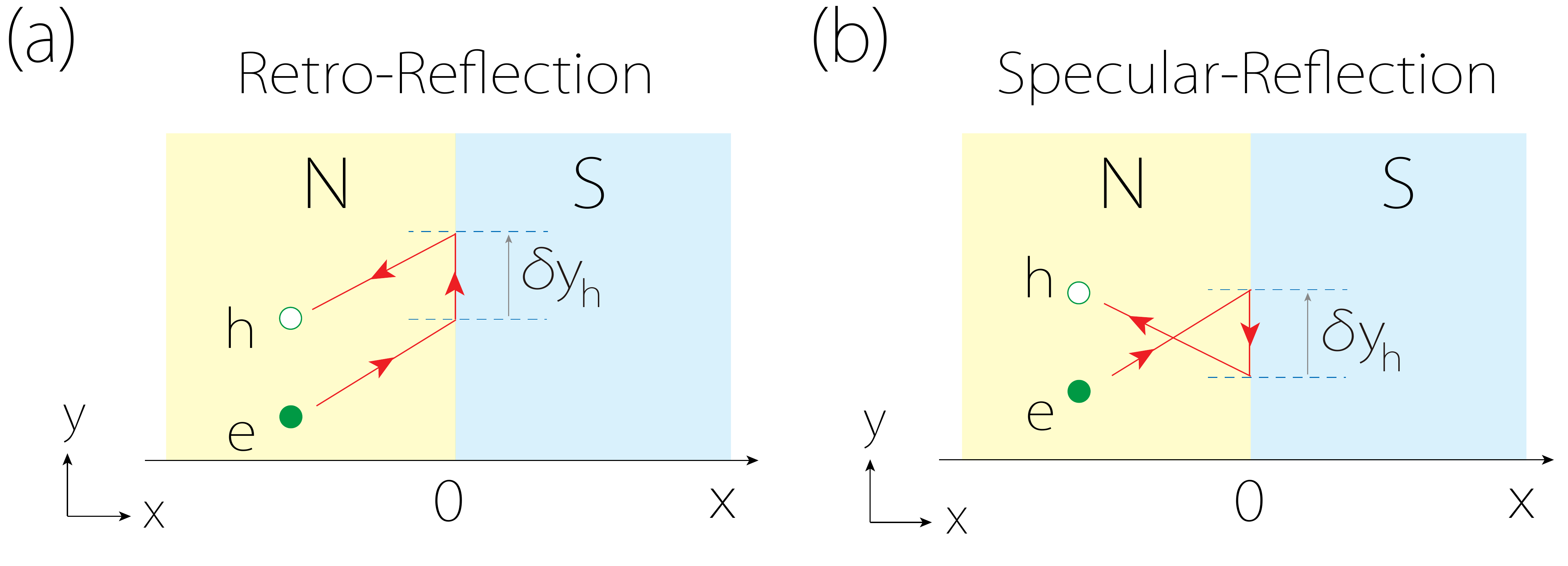}
\caption{Schematic figures for the shifts in the two types of Andreev reflection in the graphene/superconductor model. (a) is for the retro-reflection, and (b) is for the specular reflection. Note that the shifts have opposite signs for the two cases.}
\label{Fig_refl}
\end{figure}

There are also notable differences in the results between the two models.
First of all, one observes that the shift in Andreev reflection can also be negative for the graphene model [see Fig.~\ref{Fig_grasft}(d)]. This is actually connected with the specular Andreev reflection. Because of the gapless feature of the graphene band structure, the reflected hole state can be either retro-reflection from the conduction band (when $\varepsilon<E_F$) or specular reflection from the valence band (when $\varepsilon>E_F$)~\cite{Beenakker2006}. This can be understood by noticing that unlike the conduction band, the group velocity for the valence band is opposite in direction to the wave-vector.
Interestingly, from Eq.~(\ref{grasft2}), one notes that $\delta { y}_h\propto k_h^{-1} \propto {\rm sgn}(E_F-\varepsilon)$. Hence, the shift is positive for retro-reflection and negative for specular reflection [see Fig.~\ref{Fig_grasft}(f) and Fig.~\ref{Fig_refl}]. Second, the sign of $\delta y_e$ for normal reflection also depends on the relation between the different energy scales. From Eq.~(\ref{grasft1}), one observes that for $|\theta|<\theta_c$, the sign of $\delta y_e$ is determined by the sign of the factor $(E_F \tan^2\beta-\varepsilon)$. Hence, for small excitation energy, $\delta y_e$ is positive, and it becomes negative when $\varepsilon>E_F \tan^2\beta$ [see Fig.~\ref{Fig_grasft}(a,c)] [and $\delta y_e=0$ at $\varepsilon=\Delta_0$ because the factor $\tan\beta=0$ in Eq.~(\ref{grasft1})].
We also note that for $|\theta|>\theta_c$, $\delta y_e$ is pronounced and may have a sign change as a function of the incident angle.

\section{Discussion and conclusion}

In this work, we have demonstrated that sizable GH-like shifts could occur in both Andreev and normal reflections from an NS interface. In the discussion, we have focused on the sub-gap energy range with $\varepsilon<\Delta_0$. The discussion can be directly extended to the energy range with $\varepsilon >\Delta_0$. There, the shifts would vanish when $|\theta|<\theta_c$ because both amplitudes $r_{e/h}$ would then be purely real. However, the shift in normal reflection ($\delta y_e$) could be nonzero when $|\theta|>\theta_c$, due to the presence of evanescent mode for the hole reflection channel. Nevertheless, in that energy range, the probability for the reflection would be suppressed, because the quasiparticles are allowed to be transmitted into the S region [see Figs.~\ref{Fig_EG}(d) and \ref{Fig_gra}(d)].

In the discussion, we have taken a 2D system setup (as in Model I and Model II). This is because the GH-like shift is within the scattering plane. It should be clear that the effect itself is not limited to 2D systems. The analysis here directly applies for scattering at 3D NS junctions. For example, the treatment of a 3D Weyl-semimetal/superconductor junction would be very similar to the graphene/superconductor model considered here~\cite{Liu2017}, and a sizable GH-like shift would also be expected there.

\begin{figure}[tbh]
\includegraphics[width=8.4cm]{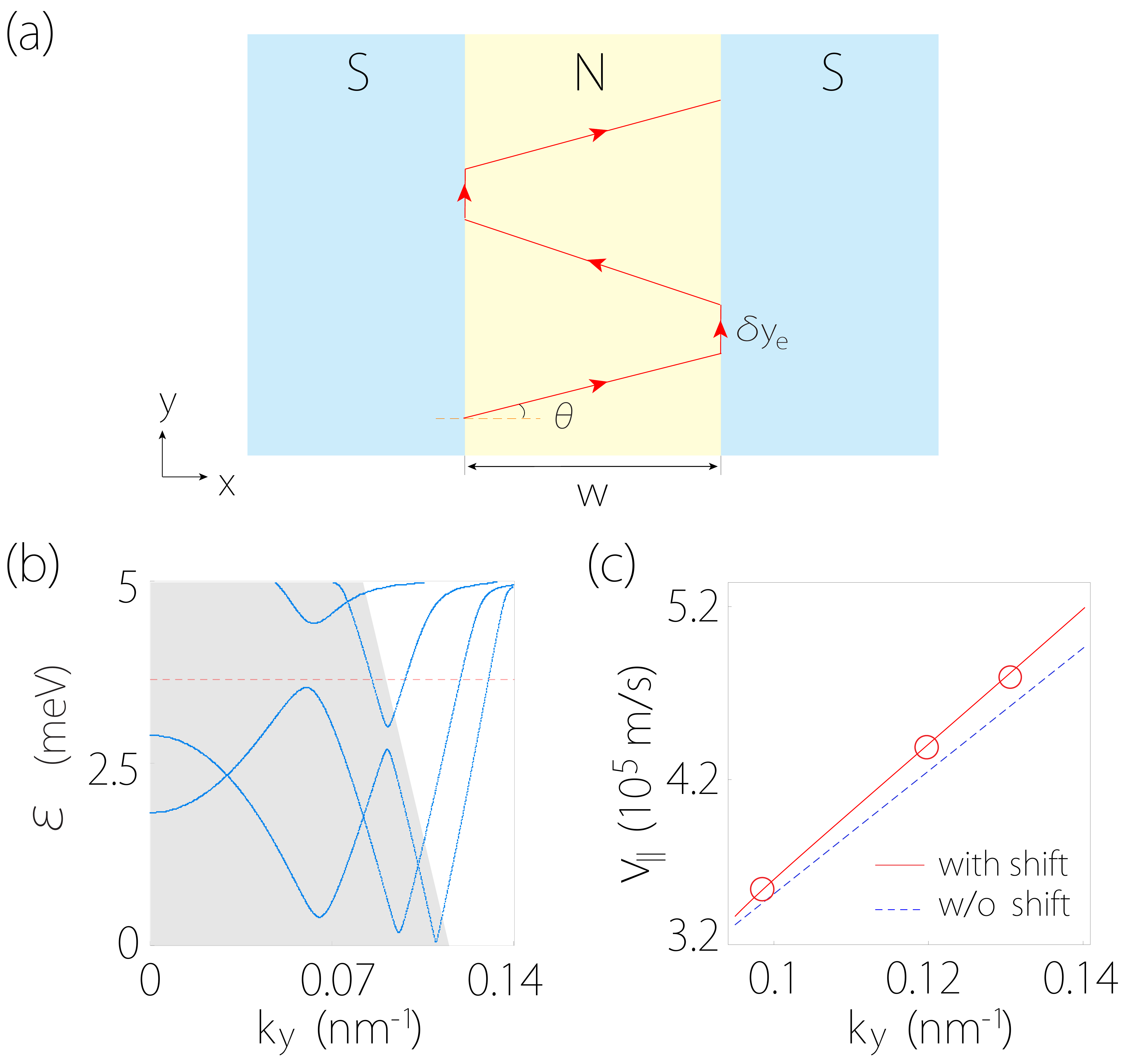}
\caption{(a) Schematic figure showing the trajectory for an electron confined in the SNS structure. Here, we assume that $|\theta|>\theta_c$ such that the electron undergoes multiple normal reflections at the two NS interfaces. Its propagation velocity along the $y$ direction is affected by the presence of the GH-like shifts. (b) Numerical results for the spectrum of the confined modes in the SNS junction. Here, we focus on the modes with $|\theta|>\theta_c$, i.e., in that unshaded region in (b). (The shaded region in (b) is for the modes with $|\theta|<\theta_c$.) (c) Group velocities $v_\|$ for the confined modes at energy $\varepsilon=3.7$ meV [marked by the dashed line in (b)]. In (c), the data points are obtained from the numerical results in (b), the red solid line is the estimation using Eq.~(\ref{vy}), and the blue dashed line is the estimation without the $\delta y_e$ correction. Here, we set $U_0=300$ meV, $E_F=10$ meV, $\Delta_0=5$ meV, $m=0.05\ m_e$, and $w=100$ nm.}
\label{Fig_Vg}
\end{figure}

The presence of the shifts have important physical consequences. We show that the shifts could strongly modify the dispersion of the confined modes in an SNS structure. The structure is schematically illustrated in Fig.~\ref{Fig_Vg}(a), which is extended along the $y$ direction. We assume that the N region with width $w$ in the middle is of a 2DEG, such that each NS interface here may be modeled as the one in Sec.~III. Assume that the two S regions have the same pair potential without any phase difference. For energies below the superconducting gap $\Delta_0$, a quasiparticle would be confined inside the N region. For example, consider an electron moving towards the NS interface with an incident angle $\theta$, when $\theta >\theta_c$, the electron would undergo repeated reflections between the two interfaces, forming a confined waveguide mode. The key point is that due to the presence of the GH-like shift in scattering, the electron trajectory is modified as in Fig.~\ref{Fig_Vg}(a), which affects the electron group velocity $v_\|$ along the $y$ direction. The effective group velocity may be estimated as
\begin{equation}\label{vy}
v_\|=\frac{\Delta \ell}{\Delta t}=v_x\Big(\tan\theta+\frac{\delta y_e}{w}\Big),
\end{equation}
where $\Delta \ell=w\tan\theta+\delta y_e$ is the distance travelled along $y$ between two subsequent reflections and $\Delta t=w/v_x$ is the time taken, and $v_x$ is the group velocity along the $x$ direction. The second term in the bracket in Eq.~(\ref{vy}) is the correction from the shift.
This estimation can be directly compared with the dispersion for the confined modes solved from the BdG equation. In Fig.~\ref{Fig_Vg}(b), we plot the spectrum obtained by numerically solving the corresponding BdG equation, where the confined electron modes for $\theta >\theta_c$ are those outside of the shaded region. In Fig.~\ref{Fig_Vg}(c) we compare the group velocities extracted from the dispersion in Fig.~\ref{Fig_Vg}(b) (the data points) and estimated using Eq.~(\ref{vy}) (the solid line). One observes a very good agreement. We also show that an obvious deviation would result, if the correction from the shift was not included [see the dashed line in Fig.~\ref{Fig_Vg}(c)].

\begin{figure}[tbh]
\includegraphics[width=8.cm]{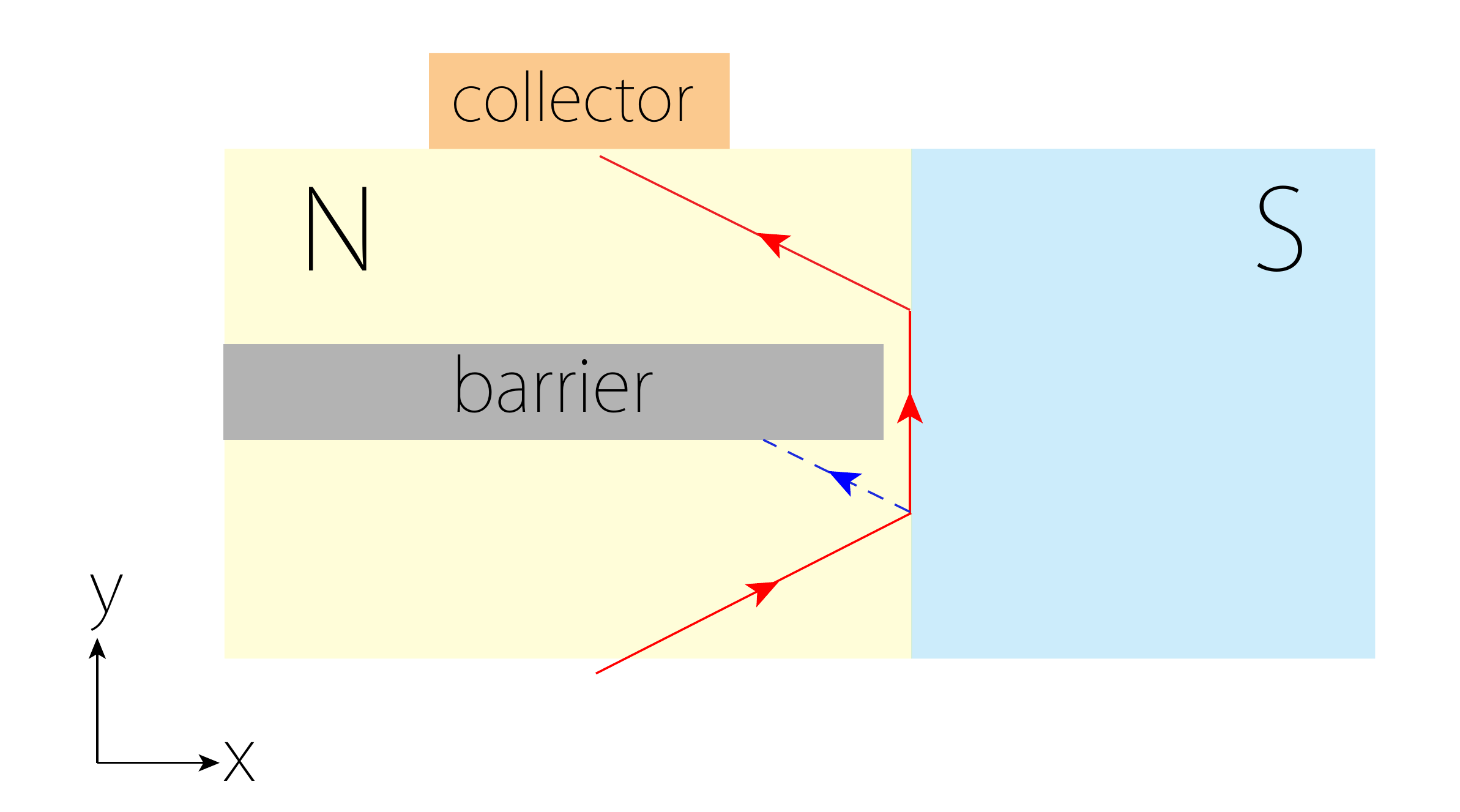}
\caption{ Schematic figure of a possible setup for detecting the GH-like shift at an NS interface. See the main text for a detailed description.
}
\label{Fig_detect}
\end{figure}

Finally, we also suggest a possible setup to detect the predicted shift in experiment. As schematically shown in Fig.~\ref{Fig_detect}, we consider an NS junction where the N side can be either the 2DEG or the graphene as discussed in this work. A collimated electron beam is incident from the N side onto the NS interface, and one tries to detect the reflected beam with the collector on the other side (see Fig.~\ref{Fig_detect}). In Fig.~\ref{Fig_detect}, we illustrate the case when the normal reflection dominates the scattering. The blue dashed line indicates the trajectory if there was no shift. We can impose a gated region (the gay colored one) such that the usual (dashed) trajectory is blocked (for graphene, this region may be an engineered hole in the sheet or another proximity-induced superconducting region). But with the anomalous shift, the beam can circumvent the barrier region and follow the red path to be detected by the collector.
By controlling the incident angle, beam energy, and the geometry of the barrier region, one can then probe the shift in experiment. To detect the shift in Andreev reflection, graphene may be a better choice, since the reflection can be tuned to be either retro-reflection or specular reflection, and one can design the corresponding geometry of the barrier region for the detection.

In conclusion, we have demonstrated the existence of sizable GH-like shifts in both Andreev and normal reflections at an NS interface. We explicitly calculated the results for two concrete systems. For the 2DEG/superconductor model, we show that the shifts become large when the N side is lightly doped. While the shift in Andreev reflection stays positive (in the forward direction), the shift in normal reflection can be either positive or negative, controllable by the quasiparticle excitation energy. For the graphene/superconductor model, we show that the shifts are more pronounced, and the sign of the shift in Andreev reflection can also be controlled, which is tied with the retro/specular reflection character. We show that as a consequence, the dispersion for the waveguide confined modes in an SNS structure is modified by the shifts. We also suggest a possible setup to probe the shifts in experiment. The discovered effect adds a new dimension for controlling the quasiparticle propagation in superconducting devices. Thus, our result not only reveals a fundamental and intriguing effect, it also opens up a new avenue for exploring potential technological applications based on NS junctions.

\begin{acknowledgements}
The authors thank Xinxing Zhou and D. L. Deng for valuable discussions. This work was supported by Singapore Ministry of Education Academic Research Fund Tier 2 (Grant No.~MOE2015-T2-2-144), NSFC under Grant No.~11534001, and NSF of Jiangsu Province, China (Grant No.~BK20160007).
\end{acknowledgements}




\bibliography{Lshift_ref}


\end{document}